# Transmission Probability in Double Quantum Well with Triple Barrier

Krishna Rana Magar, Upendra Rijal and Sanju Shrestha

Central Department of Physics, Tribhuvan University

Abstract

Quantum well of $AlGaAs/GaAs$ is very important to study transport properties of electrons due to its wider application in electronic devices. Hence, the double well of $AlGaAs/GaAs$ with triple barrier is taken to study transmission probability. Transmission probability is found to decrease with the increase in the height and width of the barrier. Transmission probability with energy of electron shows two peaks while taking all three barrier of the same height. Whereas a single and higher value of peak is found when the height of the central barrier is slightly reduced.

**Introduction**

The importance of semiconductor technology in our everyday life is being increased day by day. So, that has become a subject of great interest in terms of both theoretical and experimental studies. The theoretical interest is know the properties and characteristics of it in atomic level i.e. to say microscopic level. The way of such study is to select the appropriate material and model the material according to our requirement with latest technology such as Molecular Beam Epitaxy (MBE) and Metal Organic Chemical Vapor Deposition (MOCVD). Such modeled material structure can be used as electronics devices. Using the above mentioned techniques, devices can be made exactly of our requirement in terms of size and applications so that can be used for betterment of people.

In the mid 19[th], Leo Esaki demonstrated tunneling of electrons through a few nanometer thick barrier of semiconductor, the effect of the tunneling is used to design a diode (Esaki & Leo, 1974). In the year 1973, Esaki, Giaever and Josephson shared the very distinguished prize, the Nobel Prize in Physics for their works on quantum tunneling in solids (Esaki & Leo, 1974; Dardo, 2004).

Compound semiconductor, made by combining elements of Groups III and V such as $GaAs$, $GaP$, $InP$ etc are being used due to it tune ability of band gap when alloyed with elements such as $Al$. Hence, by taking an appropriate thickness and size of layer of $GaAs$ and $AlGaAs$, it is possible to make quantum well, quantum wire, quantum dots and even multiple quantum wells. A quantum well structure is obtained when a small band gap material like $GaAs$ is kept in between two large band gap material like $AlGaAs$. This structure is 2D as the electron is confined to move in the plane of x-y. It is equally true for the hole. Basically electrons in such structure is confined into an axis. Due to different development of epitaxial growth technology like MBE, MOCVD etc. (Ibach & Luth, 2009), as discussed above the fabrication of these structures of accurate size has become possible. In such structures, the electronic properties are drastically modified and hence, completely new and interesting properties are observed like quantum hall effect (Ibach & Luth, 2009), quantum size effect (Ando et al., 1982; Mendez, and Klitzing, 1987), resonant tunneling (Schoeller, and Schon, 2008) etc. This has opened up new arena for optoelectronics, quantum information and spintronics. Many of the physical effects in quantum well structures can be seen at room temperature and can be exploited in real devices. Hence, the authors have done some theoretical studies of the transmission probability into the double well and triple barrier structure of $GaAs/AlGaAs$. The characteristic of the transmission probability with variation of barrier width, height and the energy of electron is done. The study is considered to be important for designing new low dimensional devices.

**Theory**

The Schrodinger equations for an electron with energy $E < V$ are used:

for the well region,

$$\frac{d^2\psi_w(x)}{dx^2} + \frac{2m_w E}{\hbar^2}\psi_w(x) = 0 \tag{1a}$$

for the barrier region,

$$\frac{d^2\psi_b(x)}{dx^2} - \frac{2m_b(V-E)}{\hbar^2}\psi_b(x) = 0 \tag{1b}$$

The corresponding wave vectors are:

$k_w = \sqrt{2m_w E/\hbar^2}$, and $k_b = \sqrt{2m_b(V-E)/\hbar^2}$.

The solutions of the equations above are:

For the 'I' region i.e. outside the well,

$$\psi_1(x) = e^{ik_1 x} + re^{-ik_1 x} \tag{2}$$

$$\text{where } k_1 = \sqrt{2m_w E/\hbar^2}$$

For the 'II' region, the 1st barrier

$$\psi_2(x) = A \cosh(k_2 x) + B \sinh(k_2 x) \tag{3}$$

where $k_2 = \sqrt{2m_b(V-E)}/\hbar$

Similar equations are used for region 'IV', 'VI' with the coefficients F, K respectively in the 1st term and G, L respectively in the second term.

For the III$^{rd}$ region, the 1st well

$$\psi_3(x) = C \cos(k_3 x) + D \sin(k_3 x) \tag{4}$$

where $k_3 = \sqrt{2m(V-E)}/\hbar$

Similar equations are used for region 'V' with the coefficients H, I for 1st and the 2nd term respectively.

For the 'VII' region, outside the barrier,

$$\psi_7(x) = T e^{ik_1 x} \tag{5}$$

since there is no reflection of the wave.

Finally, the transmission probability is calculated by using appropriate boundary conditions and equality of the effective mass-weighted slope of the wave function (Ben Daniel Duke condition) at the interface (Davidson et at., 1986) and matrix method. The details of it is described by Shrestha et al. (Shrestha et al., 2011).

**Result and Discussion**

$GaAs$ is a direct and band gap material with band gap of $1.52\ eV$ and $1.43\ eV$ at $0\ K$ and $300\ K$ respectively (Vurgaftman et al., 2001). The band gap of $AlGaAs$ is more than that of $GaAs$, depending on the concentration of $Al$. It ensures better and high temperature performance with better radiation hardness i.e. the energy gap of $AlGaAs$ depends on the concentration of $Al$ in $Al_x Ga_{(1-x)} As$. It is given by $(1.42 + 1.25x)\ eV$ (Vurgaftman et al., 2001; Frensley et al., 1994) for $x < 0.45$. Double quantum well of $GaAs/AlGaAs$ and three barrier with each well width and the each barrier of $50\ Å$ is taken along with all the height of the barrier to be $0.347\ eV$ respectively, is given in the Figure 1. The energy level of electron of $0.345\ eV$ is taken as it is lower than that of the barrier. Is the condition of as $E < Vo$ (Frensley et al., 1994). The quantum mechanical transmission is studies i.e. transmission probability through the double well and three barrier is studied. Figure 2(a) shows the variation of transmission probability with height of the barrier. The probability decreases with the increase in the height of the barrier. Figure 2(b) also shows the decrease in the transmission probability with the increase in the width of the barrier. Both quantum mechanical tunneling show similar characteristics i.e. higher the height of the barrier/width the lower the transmission probability. However the transmission probability is more sensitive to height of the barrier as the probability decays rapidly with increase in the height of the barrier. The variation of the transmission probability with the energy of electron shows two peaks of energy values $0.3398\ eV$ and $0.3438\ eV$, shown in the Figure 2(c). That verifies the quantized energy states in the quantum well.

Figure 3(a) shows the variation of transmission probability with height of the barrier as in the Figure 2 (a). The different in the condition is that the height of the either barrier is taken as $0.347\ eV$ i.e. is greater than the central barrier, $0.346\ eV$. The central barrier is slightly reduced in the height. Almost similar nature of the curve is observed, however the magnitude of the tunneling probability is slighter greater than that of the earlier case. Similar results is found in the transmission probability, for the condition mentioned above versus barrier width, given in the Figure 3(b).   The variation of the transmission probability with the energy of electron, show in the Figure 3 (c) shows a single peak of energy values $0.345\ eV$. That verifies the quantized energy states in the quantum well with lower height of the central barrier as compared to that of either has higher quantized energy value is greater than that of the double well and triple barrier of the same  height.

**Conclusions and the Further work**

The transmission probability into the double well triple barrier structure decreases with increase in the height and width of the barrier for equal height of the barriers as well as for central barrier having slightly lower height in either of the two extreme barriers. Two peaks of the transmission probability of the values 0.3398 eV and 0.3438 eV are observed for double well with all the three equal height barriers. Whereas a single peak of the transmission probability value 0.3435 eV is observed for lower height of the central barrier as compared to that of the height of the either barriers.

The work can be used to design miniaturized low dimensional device to calculate appropriate height and width of the barrier to stop tunneling effect into the multiple well or in nanowire i.e. to select appropriate barrier.

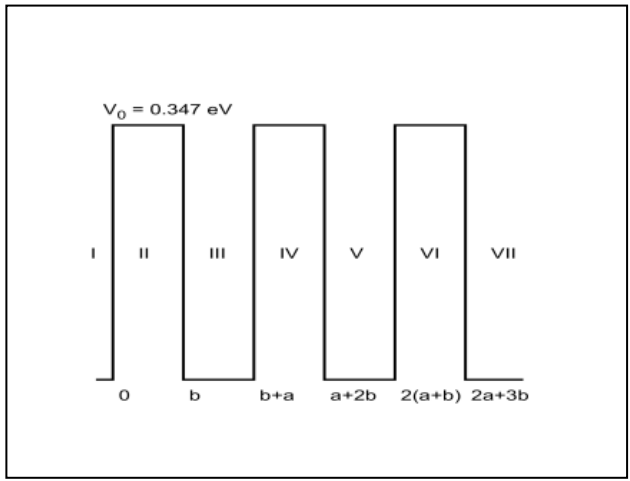

Figure (1) Double quantum well with three barrier.

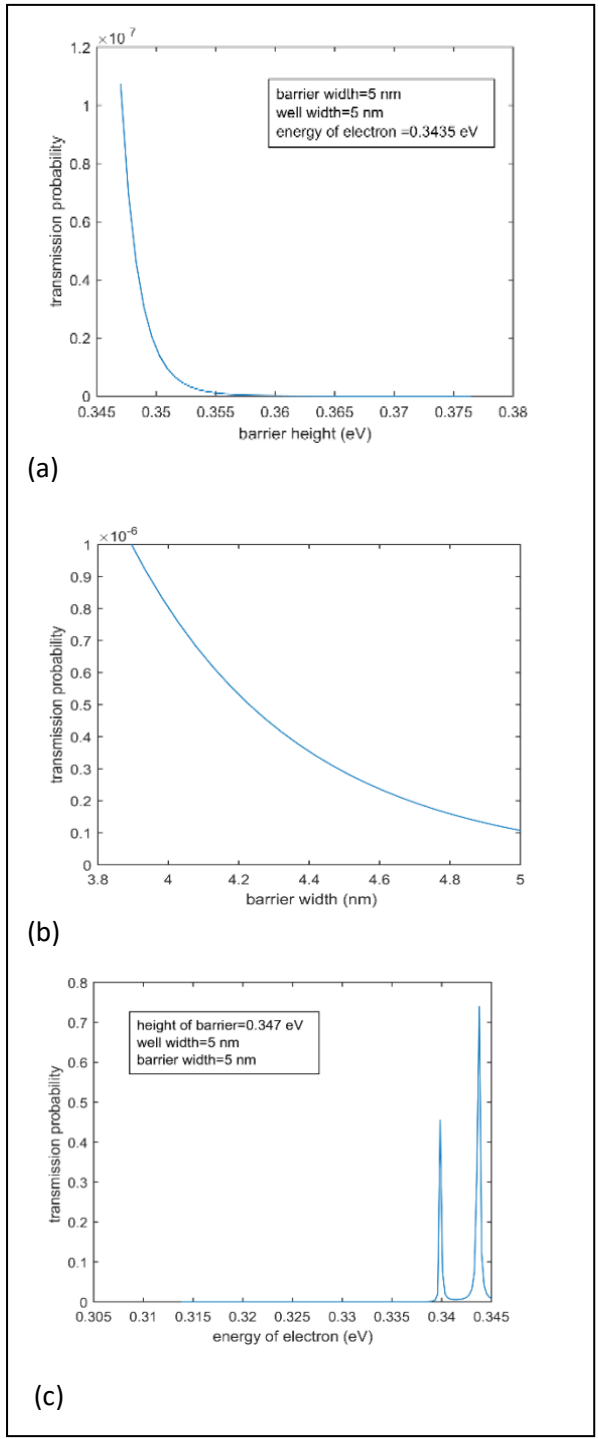

Figure (2) Transmission probability versus (a) barrier height for barrier and well width of 5 $nm$ with energy of electron 0.3435 $eV$, (b) barrier width for well width of 5 $nm$, barrier height of 0.347 $eV$ and energy of electron 0.3435 $eV$. (c) Transmission probability ($\times 10^{-7}$) versus energy of electron for barrier and well width of 5 $nm$ and height of the barrier of 0.347 $eV$.

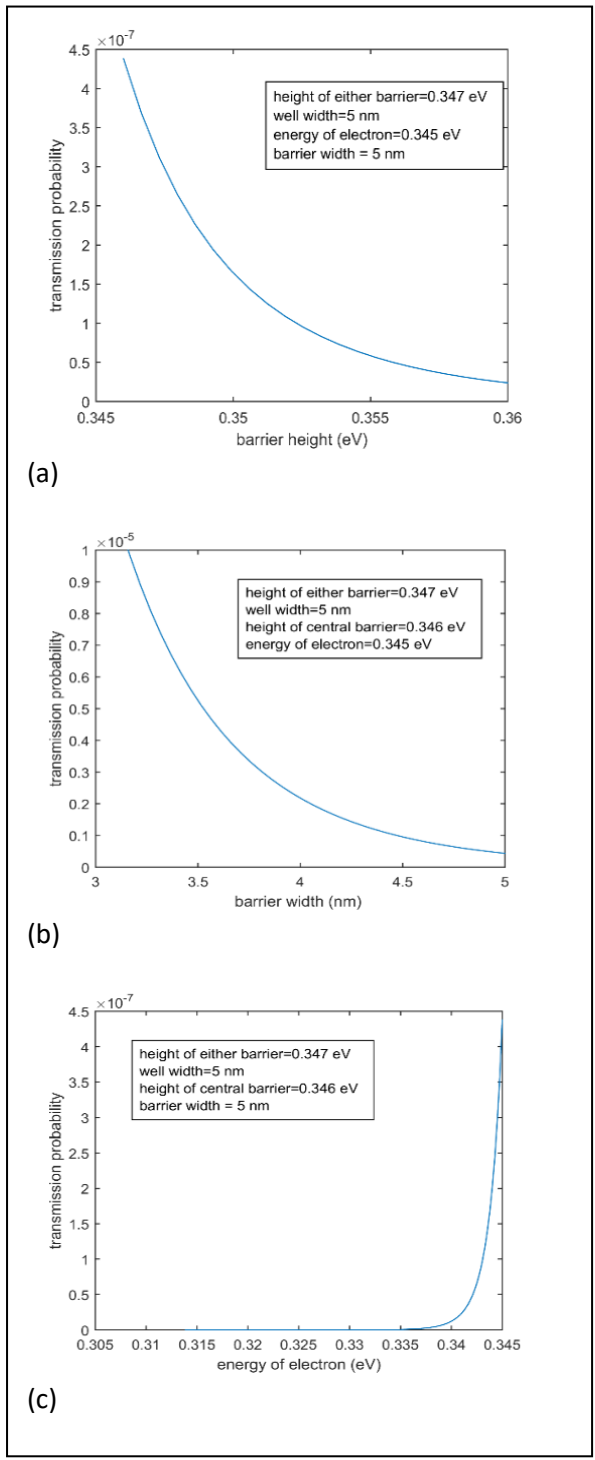

Figure (3) Transmission probability versus (a) barrier height for barrier and well width of 5nm with energy of electron $0.3435\ eV$ taking the height of the either barrier to be $0.347\ eV$, (b) barrier width for well width of $5\ nm$, barrier height of either barrier to be $0.347\ eV$, central barrier $0.345\ eV$ and energy of electron $0.3435\ eV$, (c) energy of electron for barrier and well width of 5nm and height of the either barrier of $0.347\ eV$, central barrier $0.346\ eV$.